# Observation of open Fermi surface in bismuth

Tito E. Huber[1*], Leonid Konopko[2], and Albina Nikolaeva[2]

[1] *Howard University, Washington, DC 20059, USA*
[2] *Technical University, Ghitu IIEN, Chisinau, MD-2028*

Bismuth is a candidate material for three dimensional higher dimensional topological insulators. We performed electronic transport experiments on small diameter (~50 nm) crystalline bismuth nanowires to clarify the role of the proposed hinge channels in the interlayer coupling. The magnetoresistance presents a sequence of peaks at Yamaji magic angles for which the interlayer group velocity is zero pointing to flat bands due to the layered structure. Furthermore, we observe a peak at high fields (10 T > $B$ > 14 T) for $B$ applied parallel to the layers, a definitive signature of interlayer coherence that enables deduction of the interlayer transfer integral ($t \approx 8$ meV). We demonstrate transport by a corrugated open Fermi surface of holes that is extended, with the high angular precision of ~ 0.016 (rad) in the interlayer direction. The observations indicate that the coherent and directional interlayer coupling established by hinges play a key role in the coherent interlayer coupling of bismuth.

# I. INTRODUCTION

Topological insulators (TIs) have attracted considerable interest because of their unique condensed matter properties and their applications in electronics. The counter-intuitive behaviors of TIs are rooted in the fact that they exhibit bulk insulator properties, while their surfaces are metallic and topologically protected [1]. Bismuth consists of a stack of bismuth bilayers, which, in isolation, are two-dimensional (2D) TIs [2─4]. It has been proposed that bismuth is a higher-order TI (HOTI), in which the crystal hinges host topologically protected channels [5,6]. Such modes appear when the facets of a bismuth crystal having gaps of opposite signs intersect. Previously, experimental scanning-tunneling microscope (STM) studies of bismuth (111) surfaces that have nanoscale islands addressed the role of hinges in electronic transport [1,7,8]. Hinge channels lend themselves to an unequivocal interpretation of the complex pattern of the currents observed around the islands. These studies were supplemented with innovative Superconducting Quantum Interference Device (SQUID) measurements that exploit proximity–induced superconductivity to reveal the hinge channels in crystalline bismuth nanowires surfaces. Bismuth thin films, nanowires, and ribbons, are low-dimensional forms of bismuth in which a large surface-to-volume ratio and quantum confinement favors surface electronic transport over bulk transport and that demonstrate extraordinarily high mobility [9─13] and even display quantized conductance [9,12]. They are well suited for the evaluation of the effects of topological protection. The aim of this paper is to provide experimental characterization of interlayer transport in high quality, small diameter bismuth nanowires by employing a traditional electronic transport method. In our work, we apply the method of angle-dependent magnetoresistance (AMR) to bismuth nanowires. AMR employs high magnetic fields and is very often used in the study of layered materials such as quasi two-dimensional (2D) organic conductors [14]; however, this method has never been applied in the study of bismuth nanowires. Our studies are reveal highly-directional and coherent interlayer coupling in bismuth. We find that the AMR show Yamaji [14] oscillations that are periodic in tan$\alpha$, where $\alpha$ is the angle between the magnetic field and the perpendicular to the bilayer or tilt angle. This oscillatory phenomenon was first recognized in layered quasi 2D organic conductors [14─16], it has also been observed in layered metals such as $PdCoO_2$ [17,18], and explained in terms of a corrugated tube open FS. The FS corrugation is evident in the experiments because at Yamaji magic angles $\alpha_n$ between the (111) plane and the magnetic field, the AMR has a sequence of maxima when the electron is trapped in orbits defined by the corrugation that have no energy dispersion, $v_z = dE/dk_z = 0$ or exist in a flat band. Here, $\mathbf{v} = \nabla \mathbf{E}/\hbar$ is the electron's group velocity. As the conductivity is proportional to $v_z$, the MR exhibits maxima at the magic angles [16]. Also, in addition to Yamaji oscillations, the Bi nanowires exhibit an AMR feature consisting of peaks for angles corresponding to $B$ along the layers. This is proof of coherent interlayer coupling [19─21] where the interlayer transfer integral $t$ is less than the intralayer transfer integral $t_{//}$. The identification of the FS topology, closed FS versus open FS, is key to our understanding of the metallic state [22,23] of bismuth surface states. Furthermore, the identification of flat bands in bismuth due to the FS corrugation is intriguing because flat bands promote correlated electron behavior such as superconductivity [24,25] in bilayer graphene and superconductivity has also been observed in low-dimensional bismuth [26]. The observation of an open Fermi surface, which can be interpreted in terms of hinges channels, was not anticipated as bulk bismuth crystals are characterized by electrons and holes in a closed FS with ellipsoidal pockets. These phenomena are discussed in the paper.

## II. EXPERIMENT

The nanowires were fabricated using a fiber-drawing process [12,13]. This is a significant advance in the fabrication of nanowires. Briefly, in a first step, the Ulitovsky technique was used to prepare 200-nm wires. This technique of casting at the nanoscale, involved using a high-frequency induction coil to melt a 99.999% Bi boule within a borosilicate glass capsule while simultaneously softening the glass. Glass fibers containing the 200-nm nanowires were then pulled from the capsule with a mechanical puller and annealed. X-ray and AMR show that these bismuth wires are single crystal [12]. In the second step, fibers containing 200-nm wires were stretched with a capillary puller via the Taylor method to reduce their diameter. Subsequently, the nanowires were annealed at 200 C. The resultant Bi filament was not continuous, yet sections that are a fraction of a millimeter in length could be selected using an optical microscope. Electrical connections to the nanowires were performed using $In_{0.5}Ga_{0.5}$ eutectic. This type of solder consistently makes good contacts, as compared to other low-melting-point solders, but it has the disadvantage that it diffuses at room temperature into the Bi nanowire rather quickly. Consequently, the nanowire has to be used in the low temperature experiment immediately after the solder is applied; otherwise the sequence of magnetoresistance peaks discussed here are not observed. This can be attributed to room temperature diffusion of Ga in the nanowire. The crystalline nanowires are believed to be highly faceted because they grow at an angle of $70º$ from the (111). The length of the nanowires is 0.3 mm. The temperature-dependent resistance R(*T*) saturates at an intermediate temperature because of quantum confinement and surface scattering. The diameter could be precisely characterized because the nanowires, even the very small diameter ones, exhibit Aharonov-Bohm (AB) oscillation at 4 K with an applied magnetic field along the wirelength where the period in magnetic field is inversely proportional to the square of the diameter. The magnetoresistance (MR) measured when *B* is parallel to the length of the wire (LMR) and thermopower results for Bi nanowires [13] has been examined. Results for the 50 nm sample A96 have been presented [13]. It was observed that with the increase in magnetic fields along the wire axis, the wires exhibited a stepwise increase in conductance and oscillatory thermopower. AB oscillations are caused by an increased number of high-mobility spiral (helical) surface modes for increasing magnetic fields along the wirelength and shows that the dissipation is extremely low. The mean free path of spiral modes was estimated to exceed 10 µm at 10 T.

The diameter *d* of the single-crystal Bi nanowires in the primary batch used in our experiment ranged from 45 to 55 nm. Additional batches of large diameter wires in the range of 75 nm to 340 nm were also employed in this investigation. Several experimental runs with four samples drawn from the primary batch were performed. We examined the dependency of the magnetoresistance (MR) upon the orientation of the nanowire with respect to the magnetic field (AMR) as a function of the tilt angle $\alpha$, where $\alpha$ is the angle between the magnetic field and the perpendicular of the (111) atomic plane. In these experiments the magnetic field *B*, in the range of 0–14 T, is applied perpendicular to the wire length. The geometry is shown in the inset of Fig. 1 where C3 is the perpendicular to the bilayer and C2 is the binary axis. Hereafter we shall present the AMR results for a sample (A49) of the batch. Results of our experiments with the larger diameter nanowires, are presented in Appendix B. We observed an angular dependence that is symmetric around the AMR local minimum at $\alpha = 90°$. Fig. 1(a) shows the AMR in the range of $\alpha = 20$–$100°$. The AMR had a smooth component possessing the form $(1 - \cos^2(\alpha))$, and is caused by Lorentz forces. This form of magnetoresistance has been observed in BiSb films [27] also.

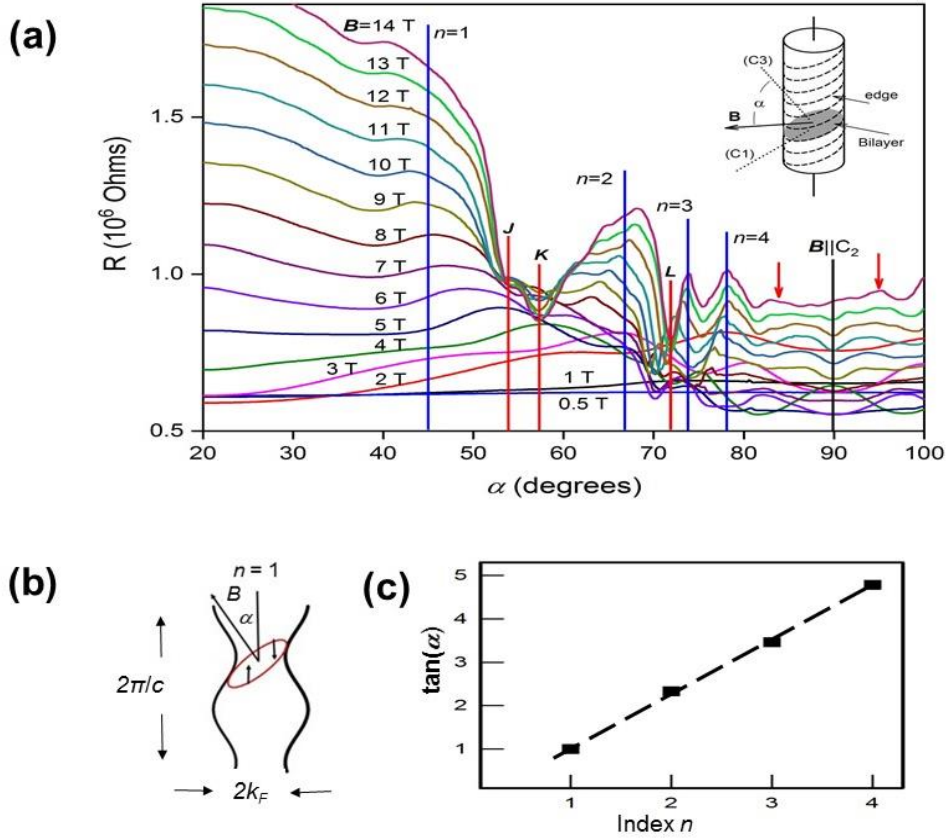

FIG. 1. Yamaji oscillations. (a) Angular dependence of the magnetoresistance (AMR) versus $\alpha$, for different magnetic fields $B$ of our 50-nm Bi nanowire sample (A49). The experimental geometry is illustrated in the inset. The spacing between bilayers is $c$. The magnetic field is perpendicular to the wire axis. The angle dependent magnetoresistance was observed experimentally by rotating $B$ around the wire axis. If $\theta$ is the angular position of $B$, the angle between the normal to the bilayer, C3, and the magnetic field is $\alpha = \theta + 20° (90° - \theta)/90°$. The resistance shows maxima at the magic angles indexed as $n$ = 1, 2, 3 and 4. The angles indicated as $J$, $K$ and $L$ are MR minima. For $J$, $\alpha = 54°$. For $K$, $\alpha = 57.4°$ and for $L$, $\alpha = 72°$. Side peaks are indicated with red arrows. (b) Illustration of corrugated tube showing bellies. The behavior of $v_z = (2\pi/h)(dE/dk_z)$ for the first Yamaji angle, as indicated. The orbit is represented with a red ellipse and the velocities are black arrows. In the latter case, the average of $v_z$ in the orbit, called drift velocity is zero, which leads to zero conductance and causes the MR peaks. (c) Observation of the linear dependence of tan($\alpha$) versus index in accordance with Eq. 1; $k_F = 7.1 \times 10^8$ m$^{-1}$.

In addition, the AMR has an oscillatory component. The oscillations are also presented by all the other samples of the batch. AMR oscillations that are periodic in $\tan\alpha$ were observed in a wide range of quasi-two dimensional materials and were successfully explained on the basis of a corrugated tube model of interlayer coupling, where the electronic states of layered materials obey the dispersion relation [14]:

$$E = E(k_x, k_y) - t \cos(ck_z) \qquad (1)$$

The first term represent the energy dispersion in the conducting layer where $k_x$ and $k_y$ are the in-plane components of $\mathbf{k}$, and $k_z$ is the component perpendicular to the plane. The second term is the interlayer coupling, where $t$ is the transfer integral and $c$ is the interlayer distance. Here, the first term in Eq. 1 represents the 2D surface states of the Bi bilayer. These states are observed via angle-resolved photoemission spectroscopy (ARPES) [26] of the (111) surface of Bi crystals as well as thin films. It was found that the surface electrons are in a $7\times10^{-2}$ Å$^{-1}$– radius ring centered on the bilayer normal axis (trigonal axis) of the 2D Fermi surface and surface holes are in six-fold ellipsoidal hole pockets with axis measured to be between 0.4 and 0.7 Å$^{-1}$. Electron and hole charges are located in a small fraction of the cross-section of the Brillouin zone perpendicular to the (111) plane. The bilayer electronic charge density $\Sigma_3$ was measured in the ARPES experiments and was estimated to be $8 \times 10^{12}$/cm$^2$. In discussing our data, we are concerned specifically with an effect of a 3D open FS that appears in a layered material such as bismuth. According to Eq. 1 the Fermi surface (FS), with $E = E_F$, is a corrugated tube, where the tube diameter has an oscillatory component with a period of $2\pi/c$ and amplitude $t$. It was shown by Yamaji [14], on the basis of a semiclassical theory, that in an open FS there are closed orbits that satisfy $ck_F \tan\alpha_n = (n - (1/4))\pi$, where $k_F$ is the projection of the Fermi wave number on the layer conduction plane, $c$ is the interlayer distance and $n$ is an integer. The angle $\alpha_n$ predicted by this equation is independent of the magnetic field in contrast to Shubnikov de Haas (SdH) oscillations [11] that show a $1/B$ dependence.

The close orbits predicted by Yamaji cause peaks of the AMR [16]. Such peaks are observed in our experiments and the phenomenon is evident in the linear relationship between $\tan\alpha_n$. The peak order $n$ is shown in Fig. 1(c). From the observed proportionality constant which is indicated with a dashed line and taking $c = 0.38$ nm [28], we find the Fermi surface tube radius is $k_F = 0.67$ Å$^{-1}$. Therefore, these Yamaji orbits must involve holes since electrons could not contribute because their momenta is limited to about 0.1 Å$^{-1}$ In light of the simulations by Yagi *et al*. [16], the width of the $n = 2$ peak of the AMR at 10 T, that is $\Delta\alpha \sim 10°$, indicates that $\omega_o\tau \approx$ 1−3, where $\omega_o = eB/m$ and $\tau$ is the relaxation time. Therefore $\tau$ is estimated to be approximately $3\times10^{11}$ sec$^{-1}$. Comparable values for the Bi surface states have been reported in the past [12,13].

Bi shows unusual interlayer coupling. For $\alpha \sim 90°$ the sample AMR is negative. A similar negative transverse MR has been observed in other layered materials. The phenomenon can be interpreted in terms of the axial anomaly [29] where $\boldsymbol{E}$ is aligned with $\boldsymbol{B}$. The emergence of the axial anomaly can be tied to the topological properties of the surface states. Pippard reviewed similar non-saturating effects associated with an open FS including the "whiskers" of the AMR in some metals like copper caused by the growth of the MR without saturation [30]. In close correspondence, bismuth nanowires also show non-saturating MR peaks for the tilt angles that we identify as Yamaji's. This data is presented in Appendix A.

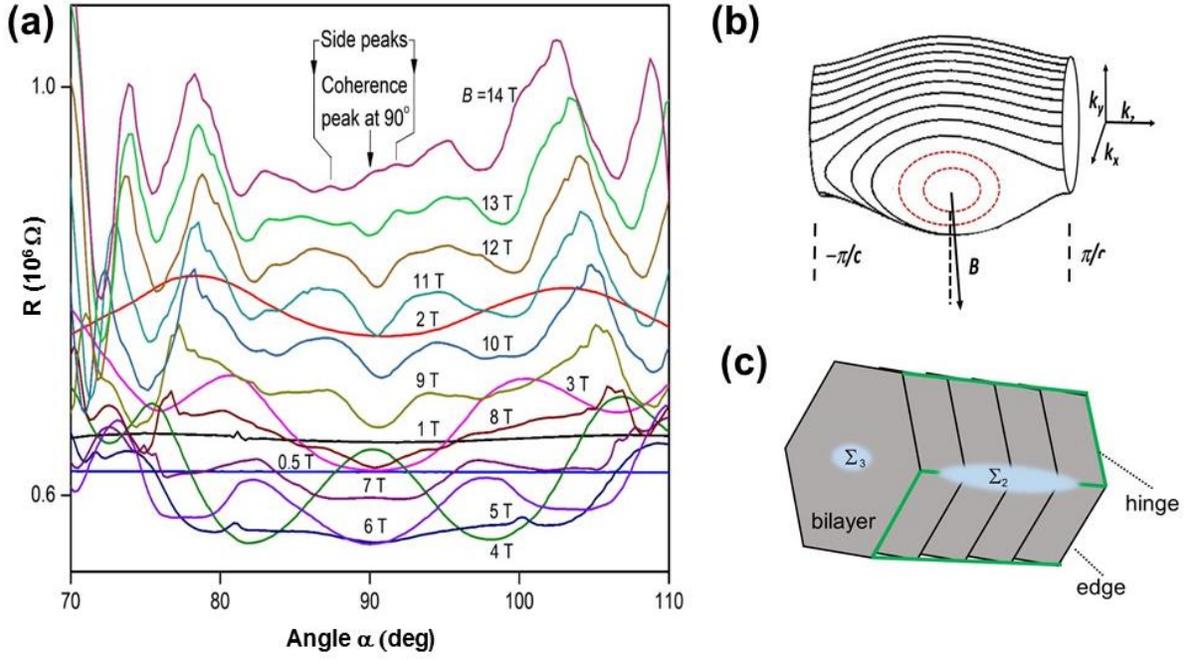

FIG. 2. Coherence peak. (a) AMR of the 50-nm nanowire showing the coherence peak and sidebands for $\alpha \sim 90°$. The nanowire shows negative magnetoresistance in a range of from 75° to 115° for intermediate magnetic fields (3 T $< B <$ 8 T) (b) Schematic diagram of the corrugated tube Fermi surface of the bismuth surface in $k$-space. The radius of the cylinder is $k_F$. Vertical dashed lines show the Brillouin zone boundaries at $\pm\pi/c$. Open orbits are represented with solid black lines and the closed orbits, that give rise to the $90^0$ peak (Ref. 20) are represented with dashed red lines. Proposed schematic of the layer structure . The bilayers are illustrated as gray hexagons. Edge and hinge modes are illustrated with black and green lines, respectively. $\Sigma_3$ is the surface charge on the bilayer and $\Sigma_2$ represents the excess surface charge that was observed via SdH, Huber *et al*. (Ref. 13).

Further information about the characteristics of Bi interlayer coupling was obtained by examining the AMR of the A49 sample for $\alpha$ between 70 and 110° (Fig. 2(a)). This sample is 50 nm in diameter. Other experiments with larger diameter nanowires are discussed in Appendix B. The MR is a local minimum for $\alpha = 90°$ because the Lorentz forces are minimal for that tilt angle. However, at high magnetic fields, we observe a feature that includes a sharp MR peak at $\alpha = 90°$ and side peaks, located at 87° and 92°. The pattern of 90º peak and side peaks is well known in the field of layered materials [19,20]. The side peaks, which are termed kinks by Hanasaki *et al* [20] are more robust than the peak at 90º in our measurements. The coherence peak and side peaks are attributed to coherent electron transport along small closed orbits on the bulges of a corrugated cylindrical FS and also to open trajectories lying near self-crossing orbits [21]. Such orbits, which are illustrated in Fig. 2(b), are effective at increasing dissipation and lead to an increase in the MR for $α \sim 90°$. If $B$ is tilted away from the in-plane direction by an angle $Δα \sim tck_F/E_F$, such that the small closed orbits about the bulges cease to be possible, then according to these results the magnetoresistance should decrease. In our nanowires, the full-width at half maximum (FWHM) of the peak was 6°. Assuming the values of the mass $m_3 = 0.22$ from ARPES [26] and using $E_F = \hbar^2(k_F)^2/2m_3$ we find $E_F = 76$ meV. Therefore, we estimate that $t = 8$ meV. In comparison, the in-plane nearest neighbour hopping energy $E_i$ is 1.4 eV [28]. The inequality $t << E_i$ indicates that the electrons scatter more frequently than they tunnel between layers and therefore it makes sense to consider a FS that extends in the interlayer direction. Also, the surface of bismuth is found to be very anisotropic ($t << E_i$) and the small value of $t$ is the reason that STM measurements [3,4] show that the bilayer is isolated from the bulk bismuth substrate. Clearly, the FS is found to be aligned with the (111) direction, perpendicular to the bilayers, parallel to the hinges with a $Δα$, that is only a few degrees. The side peaks, a common occurrence in the observation of the 90º peak, are attributed to dissipative processes in the closed orbit [21] and their observation supports the observation of the coherence peak. It is important to note that the value of $Δα$ that we observe is typically a factor of four larger than in other cases of coherence peaks. For example, in the case of $PdCoO_2$, Kikukawa reports that $Δα \sim 1.5º$. This is because $k_F$ is especially small for bismuth compared with the other layered materials that have been studied in the past.

Now, we will discuss further effects in the data presented in Fig. 2(a). The oscillations observed between 80° and 100° for increasing *B* are periodic in 1/*B*, and can be interpreted as SdH oscillations owing to the filling and emptying of Landau levels (LL); therefore, they cannot be interpreted in terms of the semiclassical model presented by Yamaji [14] and Yagi [16]. We have observed these oscillations in other small diameter samples and reported about them in a prior publication [13]. We assigned these SdH oscillations to the 2D LL of the surface carriers present on the nanowire surface, which have orbits perpendicular to the binary axis. Analysis of the temperature and magnetic field dependence of the SdH oscillations in our prior work showed that $m_2 = 0.25 \pm 0.03$. The charge density per unit area $\Sigma_2$ was estimated from the SdH period ($P = 0.060\ T^{-1}$) using $\Sigma_2 = f/(Ph/e)$, where *f* is the 2D Landau level degeneracy, which is two on account of the two-fold spin degeneracy, and where *h/e* is the quantum flux. We found that $\Sigma_2 = 8.1 \times 10^{11}/cm^2$, which is substantial, just an order of magnitude smaller than $\Sigma_3$, the bilayer charge density [28]. Also, the SdH method has been applied to the measurement of surface states by Ning *et al* [11]. They report a low value of surface charge smaller than $\Sigma_3$, and negative MR. Observation of $\Sigma_2$ is very significant since the three-dimensional TI or strong TI (STI) would manifest itself with protected surface states at all surfaces of bismuth, including those perpendicular to the layers. Ning et al also report about the non-trivial properties of the surface states. However in our experiment, analysis of the data leads to a determination of the Berry phase *γ* that has significant errors ($\Delta\gamma \approx 0.2\ \pi$) and we cannot arrive to a conclusion regarding the topological nature of $\Sigma_2$.

### III. DISCUSSION

Our observations of an open Fermi surface suggest a commentary about the hinges. It is encouraging that HOTI predicts a highly directional transmission of longitudinal modes at the surface and that we observe a tubular open Fermi surface that correspond to this expectation. It is observed that the tube has a diameter determined by $k_F$ and therefore, there is back-and-forth exchange of momentum and energy, or strong hybridization, tending to an equilibration between hinge modes and the intralayer modes that is, the 2D modes. This is a mechanism that should be included in HOTI in order to interpret the FS tube diameter.

In sharp contrast to the observation of open FS, the bulk bismuth Fermi surface consists of ellipsoidal, closed, electron, and hole pockets [31]. AMR studies have been performed and it has been observed that the MR of bulk bismuth exhibits the angular dependency that semi-classic transport theory predicts for a multi-valley system with anisotropic mobility [32]. In these experiments, Yamaji oscillations are not observed. This confirms that the origin of the Yamaji angles and the corresponding flat bands is the bismuth hole surface states that populate the nanowires, with the underlying cause being the layered structure of bismuth. In comparison, the well-known magic angles in twisted bilayer graphene appear because of the moiré pattern in the atomic positions of carbon atoms, a geometric property that is engineered by twisting that leads to flat bands [35]. To our knowledge, there is no analog in bismuth nanowires. However, since higher-order topology have been recently extended to include new families of layered compounds, such as BiI, $MoTe_2$ and $WTe_2$ [36-38], our discovery of an open Fermi surface in bismuth creates approaches for the observation of flat bands (Fig. 1), an axial anomaly (Fig. 2), and superconductivity in the novel higher-order topological candidates. Hofmann has discussed the observation of superconductivity in Bi nanoclusters [26]. New properties and new applications may follow. In a recent and exciting proposal, crystalline Bi nanowires provide a platform for realizing Majorana modes for quantum computing [39].

In conclusion recent theoretical approaches, HOTI and TCI, consider bismuth to be a stack of bilayers, joined by van der Waals forces and predict that bismuth interlayer electrical coupling in the stack involve topologically protected one-dimensional edge and hinge states. We investigated electronic transport in small-diameter single-crystal bismuth nanowires. We found strong evidence of transport between bilayers, indicating coherent electronic transport via an open Fermi surface—a corrugated tube. The FS is found to be aligned with the (111) direction, perpendicular to the bilayers, parallel to the hinges. Therefore, we postulate that the strong coherent and directional interlayer coupling established by hinges plays a major role in the coherent interlayer coupling of bismuth. Also, our study reveals that bismuth presents an unique opportunity for studying open Fermi surfaces since it has flat bands that give rise to a sequence of magnetoresistance peaks at Yamaji magic angles. Also, we uncovered remarkable similarities between bismuth and the traditional layered materials.

## ACKNOWLEDGEMENTS

This work was supported by the project ANCD 20.80009.5007.02 in Moldova. In the U.S., the work was sponsored by the U.S. National Science Foundation STC Center for Integrated Quantum Materials, Grant 1231319, The Boeing Company, and the Keck Foundation.

**APPENDIX A: MORE INFORMATION ON THE AMR NEAR THE YAMAJI ANGLES**

In this section we discuss, further evidence of Yamaji angles that consists of the observation of an interplay of saturated and unsaturated MR. This is shown in Fig. 3 Unsaturated MR is a result of open electron trajectories, which require a description in terms of an extended open FS. Figure 4 shows the nanowire resistance for various Yamaji magic angles maxima as well as three minima labeled *J*, *K* and *L*, also seen in Fig. 1. It is noted that the magnetoresistance does not saturate for the tilt angles corresponding to the Yamaji angles, $n = 1$ to 4, that correspond to MR maxima. In contrast, for the minima, *J*, *F*, *K* the resistance MR saturates at low magnetic fields (2 T). This property that is reminiscent of the effect of unsaturated MR, or "whiskers", in metals, has been reviewed by Ziman [23] and by Pippard [30]. These effects in bulk copper have been reviewed by Klauder *et al* [41].

It has been shown by Yagi *et al.* [16], that this effect is a property of a conductor with a tube Fermi surface. He presented a study of the MR of a conductor with such a corrugated FS, based on the calculation within the semiclassical approximation, of the Boltzmann equation using Shockley tube integral. He found that this effect is reproduced in the calculations for various $\omega\tau$, including those in the range of 1−3 that is relevant here.

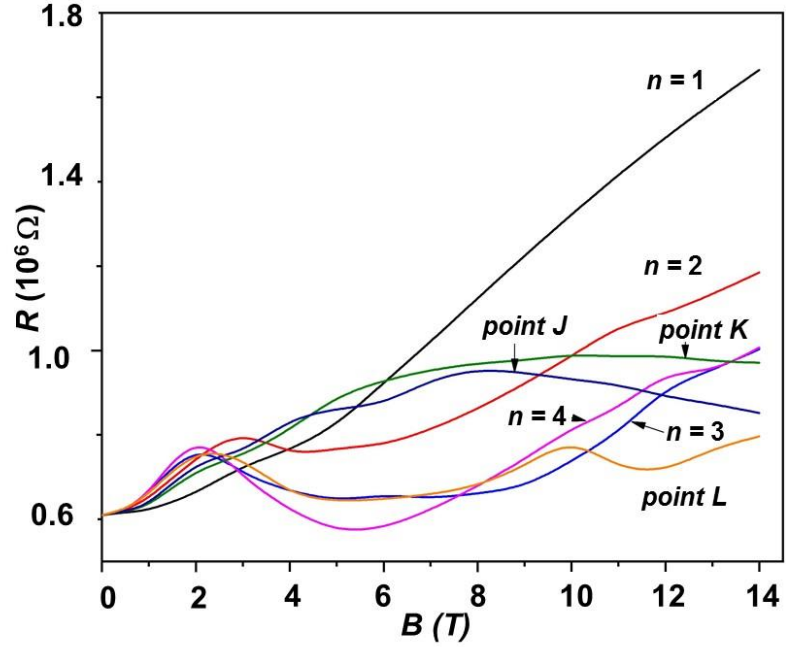

Fig. 3. Nanowire resistance as a function of *B* for various AMR maxima and minima several angles, that is Yamagi angles of order $n = 1$ to $n = 4$, and for AMR minima *J*, *F*, and *K*.

**APPENDIX B. COHERENCE IN LARGE DIAMETER Bi NANOWIRES**

In this appendix, we present data of large diameter nanowires. The fiber-drawing process [12,13] is a significant advance in the fabrication of Bi nanowires. It allowed us to fabricate nanowires of various diameters. The diameter is estimated using the Aharonov-Bohm oscillation of the MR under an applied field B along the wirelength. Nanowires in the range of diameters *d* between 90 nm and 210 nm show the coherence feature, consisting of a coherence peak and side peaks at 90°. These cases are in addition to the case of 50-nm nanowires shown in Figs. 1 and 2. In contrast, nanowires of diameter 340 nm and above, show a minimum at 90° and no coherence feature. This is because the bulk electron and holes dominate the transport and their FS is closed.

The identification of the FS topology, closed FS versus open FS, is key to our understanding of the metallic state in the surface states of Bi. It has been shown that a three-dimensional Fermi surface, such as shown in Fig. 1(b) is not necessary for the observation of the Yamaji oscillations shown in Fig. 1(a). Perez Moses *et al* presents a review and discussion of the experimental proof of open FS [40]. Based on this discussion, we focus on the observation of the 90º coherence peak and side peaks.

The data is presented in Fig. 4. The side peaks at 84º and 95º are especially prominent in the 140-nm nanowire. In contrast, the 340 nm nanowire show a parabolic minimum and no coherence maximum or side peaks. The trend towards a 90º minimum is observed in the 260 nm nanowires, also. From these observations we infer that the local maximum at 90º can be linked to the dominance of surface states in electronic transport of small diameter nanowires. We also find that the width of the coherence maximum is independent of the wire diameter indicating that it is an intrinsic property of bismuth.

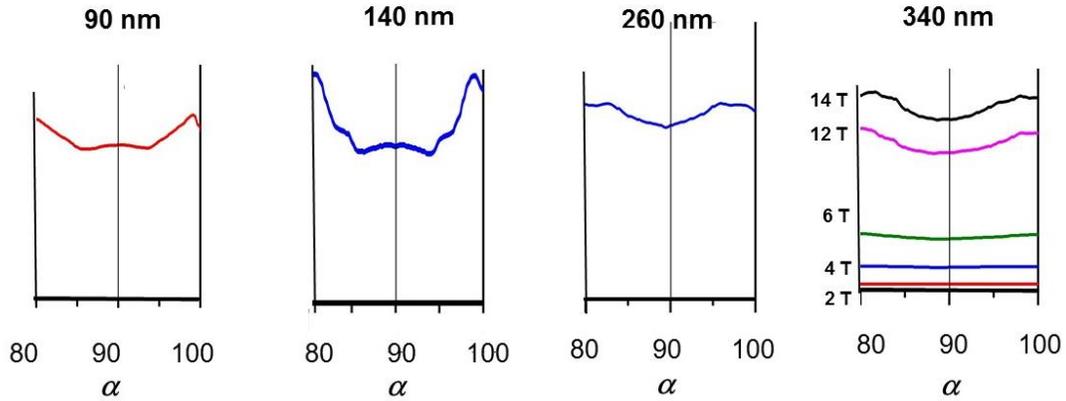

Figure 4. Coherence feature in Bi nanowires. AMR of Bi nanowires in the range 80-100 degrees of samples with four representative diameters from 90 nm to 340 nm. The coherence peak is the 90º maximum that is observed in the 90 nm and 140 nm nanowires. Among the various nanowires, the side peaks are most prominent in the 140 nm bismuth nanowires.